# Dust modelling and a dynamical study of comet 41P/Tuttle-Giacobini-Kresak during its 2017 perihelion passage

F. J. Pozuelos[1]*, E. Jehin[1], Y. Moulane[1, 2], C. Opitom[3], J. Manfroid[1], Z. Benkhaldoun[2], and M. Gillon[1]

[1] Space sciences, Technologies and Astrophysics Research (STAR) Institute, Université de Liège, 19C Allée du 6 Août, B-4000 Liège, Belgium.  
e-mail: fjpozuelos@uliege.be

[2] Oukaimeden Observatory, High Energy Physics and Astrophysics Laboratory, Cadi Ayyad University, Marrakech, Morocco.

[3] European Southern Observatory, Alonso de Cordova 3107, Vitacura, Santiago, Chile.

Accepted

## ABSTRACT

*Context.* Thanks to the Rosetta mission, our understanding of comets has greatly improved. A very good opportunity to apply this knowledge appeared in early 2017 with the appearance of the Jupiter family comet 41P/Tuttle-Giacobini-Kresak. The comet was only 0.15 au from the Earth as it passed through perihelion on April 12, 2017. We performed an observational campaign with the TRAPPIST telescopes that covered almost the entire period of time when the comet was active.  
*Aims.* In this work we present a comprehensive study of the evolution of the dust environment of 41P based on observational data from January to July, 2017. In addition, we performed numerical simulations to constrain its origin and dynamical nature.  
*Methods.* To model the observational data set we used a Monte Carlo dust tail model, which allowed us to derive the dust parameters that best describe its dust environment as a function of heliocentric distance: its dust production rate, the size distribution and ejection velocities of the dust particles, and its emission pattern. In order to study its dynamical evolution, we completed several experiments to evaluate the degree of stability of its orbit, its life time in its current region close to Earth, and its future behaviour.  
*Results.* From the dust analysis, we found that comet 41P is a dust-poor comet compared to other comets of the same family, with a complex emission pattern that shifted from full isotropic to anisotropic ejection sometime during February 24-March 14 in 2017, and then from anisotropic to full isotropic again between June 7-28. During the anisotropic period, the emission was controlled by two strongly active areas, where one was located in the southern and one in the northern hemisphere of the nucleus. The total dust mass loss is estimated to be ~ $7.5 \times 10^8$ kg. From the dynamical simulations we estimate that ~3600 yr is the period of time during which 41P will remain in a similar orbit. Taking into account the total mass loss per orbit, after 3600 yr, the nucleus may lose about 30% of its mass. However, based on its observed dust-to-water mass ratio and its propensity to outbursts, the lifetime of this comet could be much shorter.

## 1. Introduction

Minor bodies, especially comets, are invaluable sources of information that allow us to better understand how the solar system formed. They are considered to be time capsules and planetary building blocks, and are the oldest and least-evolved bodies left over from the primitive proto-solar system. They therefore represent the earliest record of material from this epoch. In addition, they are the most organic-rich bodies in the proto-solar system, and the fully formed molecules contained within their nuclei could have played a key role in the origin of life on Earth. Moreover, it is thought that minor bodies also played an important role in the hydration process of early Earth; see e.g. Hartogh et al. (2011) and Jewitt et al. (2007). Our understanding and knowledge of comets have been revolutionised by the outstanding Rosetta mission (Taylor et al. 2015) to the Jupiter family comet (JFC) 67P/Churyumov-Gerasimenko (hereafter 67P). Snodgrass et al. (2017) concluded from the observing campaign of 67P that Rosetta was seeing a typical JFC object. This allows the conclusions from Rosetta measurements to be taken as generally true for JFCs. During April 2017, the close encounter between 41P/Tuttle-Giacobini-Kresak (hereafter 41P) and Earth offered a unique opportunity to apply the lessons learned from Rosetta to other JFCs. To date, comet 41P has been poorly studied, where the only published literature was the work of Kresak (1974) who reported on two large outbursts suffered by the comet in 1973. However, 41P came to international attention because of its surprisingly fast rotational variation reported by Bodewits et al. (2018) as gleaned from observations obtained from March to May, 2017. As well as being discovered in 1858 by Horace Parnell Tuttle at Cambridge, 41P was also independently identified by Michael Giacobini in 1907 in Nice. However, it was not until 1951 when Lubor Kresák performed a third, independent identification at Skalnaté Pleso, that enough information was obtained for a proper orbit characterisation. The comet was classified as a member of the JFCs, which are assumed to come from the trans-Neptunian region, where the comets are dynamically controlled by Neptune, and in some cases are injected inwards passing to the status of Centaurs, the direct progenitors of JFCs. When Centaurs finally fall under the gravitational control of Jupiter, they become JFCs (Duncan et al. 1988; Levison & Duncan 1997). On the other hand, it has been suggested that JFCs could have other source regions closer to the Sun, such as the main asteroid belt (Fernández & Sosa 2015), the Hilda family in the 3:2 Mean Motion Resonance with Jupiter (Di Sisto et al. 2005; Toth 2006), or, with an extremely low contribution, Jupiter's Trojans (Volk & Malhotra 2008).

In this paper we present the results of a dust analysis using observations obtained with a large monitoring campaign carried out with the TRAPPIST telescopes (Jehin et al. 2011). In a second paper, an analysis of the gas content of the coma will be presented by Moulane et al. (2018). The observations were obtained between January and July, 2017, which encompasses both pre- and post-perihelion epochs. In the first part of this paper, we determine the evolution of the dust environment of 41P throughout the time the comet was active using Monte Carlo dust tail simulations (Moreno et al. 2012). This provides us information such as the dust production rate, the size distribution and ejection velocities of the dust particles, and the emission pattern. We compare our results with those obtained by other authors for similar comets, especially with those obtained from the Rosetta mission; for example, Rotundi et al. (2015), Fulle et al. (2016b), and Moreno et al. (2017a). In the second part of the study, we analyse the comet's orbital stability using numerical simulations. We also characterize its dynamical nature, which allows us to constrain its origin and future evolution. By merging our dust analysis results with its dynamical characteristics, we have developed a better understanding of 41P's current and future behaviour.

# 2. Observations and data reduction

We performed long-term, high-cadence monitoring of comet 41P using the TRAPPIST network, that is, TRAPPIST-South at ESO-La Silla Observatory in Chile and TRAPPIST-North at Oukaimeden Observatory in Morocco at several epochs pre- and post-perihelion. Both are 60 cm Ritchey-Chretien telescopes, which have thermoelectrically cooled detectors: a $2K \times 2K$ FLI Proline CCD camera with a field of view of $22' \times 22'$ at TRAPPIST-South (Jehin et al. 2011) and an Andor IKONL BEX2DD CCD camera at TRAPPIST-North. For our dust modelling purposes, we used the broad-band $R$ Johnson-Cousins filter to minimise gas contamination due to intense emission bands located in the UV and the blue portion of the comet's spectrum.

Our observational campaign started on January 20, 2017 ($r_h$=1.49 au, inbound), when the comet was bright enough to be observed with TRAPPIST-South. We monitored the comet, mostly with TRAPPIST-North, until it was too faint to be detected on July 27, 2017 ($r_h$=1.69 au, outbound). During that period, we obtained more than 40 nights of observations. The final data set is composed of 30 photometric nights, as we discarded observations obtained on cloudy nights. Our observational log is shown in Table 1.

In order to improve the signal-to-noise ratio (S/N), the comet was imaged several times each night using integration times in the range 60-120 s. The individual images were flat-fielded and bias subtracted using standard techniques, then a median stack was obtained from the available images. The flux calibration was done using the USNO-B1 star catalogue (Monet et al. 2003), so that each image we acquired was calibrated in mag arcsec$^{-2}$, and then converted to Solar Disk Units (SDUs). We oriented each image so that north is up and east is to the left. When necessary, the images were rebinned to have physical dimensions small enough as to be analysed with the Monte Carlo dust tail code. The complete set of observations is displayed in Fig. 1.

# 3. Dust analysis

## 3.1. The Monte Carlo dust tail model

The dust analysis was performed using the Monte Carlo dust tail code described in Moreno et al. (2012). The code computes the brightness of a cometary tail by generating synthetic images that can be directly compared with observations. The code has been extensively used to characterise the dust environments of comets as a function of the heliocentric distance; see, for example, Moreno et al. (2014b), Pozuelos et al. (2014b, 2015), and activated asteroids; see, for example, Moreno et al. (2014a, 2016b, 2017b).

The model computes the trajectory and the scattering of a large number of particles ejected from the object's surface. The particles, after being ejected from the nucleus, are subjected to the gravitational force of the Sun and radiation pressure. The gravity of the comet itself is neglected, which is a valid approximation for small size objects. Any gas molecules, arising from sublimated ice, drag and accelerate the dust particles to their terminal velocities. These velocities and their physical properties are responsible for the final Keplerian motion around the Sun. The ratio of the radiation pressure to the gravity exerted on each particle is given by the $\beta$ parameter, which is expressed as (Finson & Probstein 1968):

$$\beta = \frac{C_{pr}Q_{pr}}{2\rho r},$$

(1)

in this equation, $C_{pr}$ is given by:

$$C_{pr} = \frac{3E_\odot}{8\pi c G M_\odot},$$

(2)

where $E_\odot$ is the mean solar radiation, $c$ is the light speed, $G$ is the gravitational constant, and $M_\odot$ is the solar mass. $Q_{pr}$ is the radiation pressure coefficient, which is ~ 1 for particles of radius $r \geq 1$ $\mu$m (Moreno et al. (2012), their Fig. 5), $\rho$ is the particle density, and $r$ is the radius.

Since the model has many parameters, a number of assumptions were made to make the problem more tractable. With this aim, we followed the recent application of the Monte Carlo dust model for 67P by Moreno et al. (2017a), where the authors developed a detailed dust analysis based on their large ground-based observational data set and a useful compilation of the results of the Rosetta mission regarding the properties of its dust particles. Here we briefly summarise these assumptions, and we refer the reader to Moreno et al. (2017a) and the references therein for further information.

From the Grain Impact Analyser and Dust Accumulator (GIADA; Colangeli et al. (2007)) and the Optical, Spectroscopic, and Infrared Remote Imaging System (OSIRIS; Keller et al. (2007)) measurements, the density of the particles and the geometric albedo were set to values of $\rho = 800$ kg m$^{-3}$ and $p_v = 0.065$, respectively (Fulle et al. (2016b), Fornasier et al. (2015)). From the Micro-Imaging Dust Analysis System (MIDAS; Riedler et al. (2007)), we assumed that the minimum size of the particles is 10 $\mu$m, which is time-independent. On the other hand, the maximum size of particles is considered time-dependent ranging from 1 cm at large heliocentric distances (Rotundi et al. 2015) to decimeter-sized aggregates during the perihelion passage (Fulle et al. 2016a). The size distribution was assumed to be a power-law function, $n(r) \propto r^{\delta(t)}$, where the time-dependent parameter $\delta(t)$ is set to vary between -4.2 and

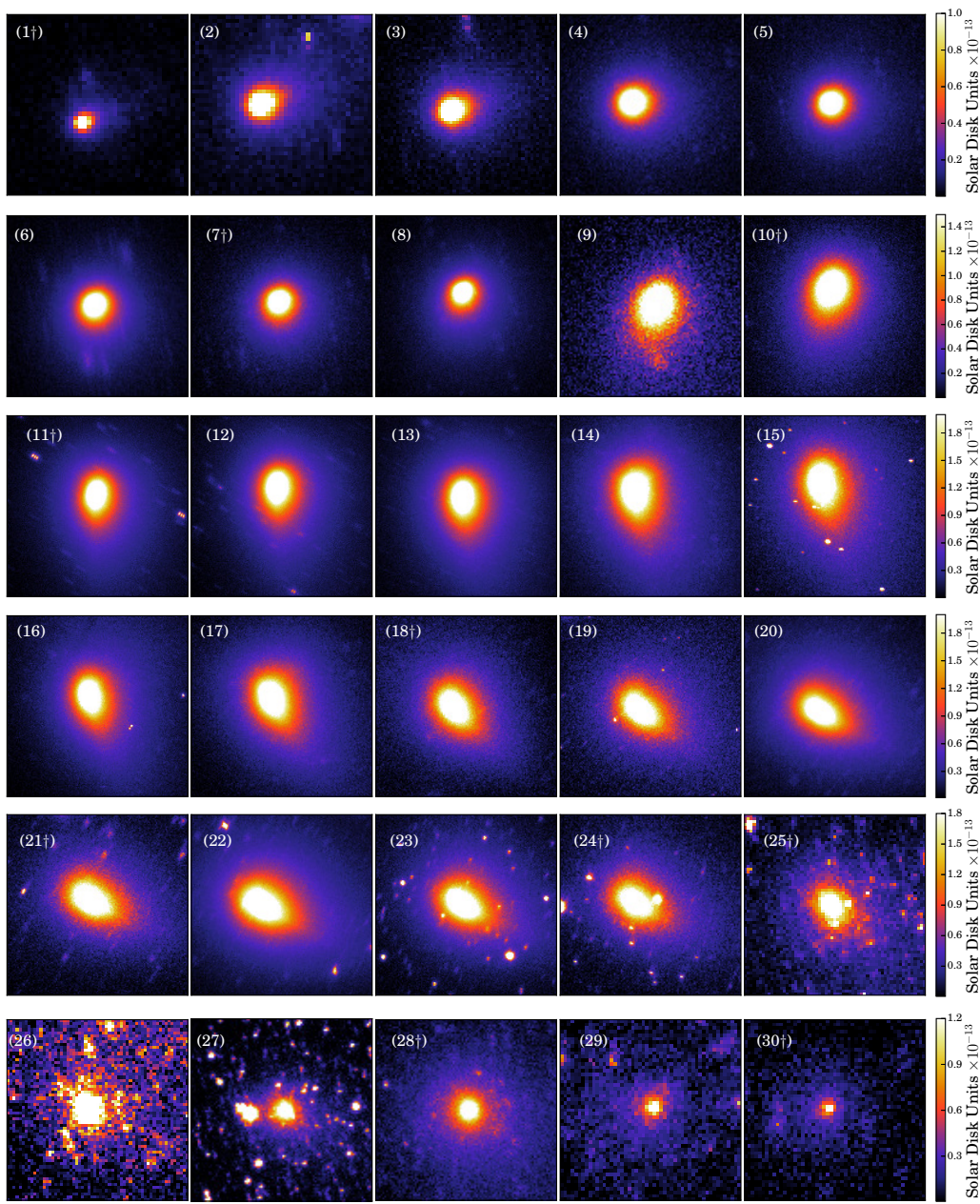

Fig. 1: Complete set of observations using both TRAPPIST-South and TRAPPIST-North telescopes at La Silla Observatory and Oukaimeden Observatory, respectively. Isophote levels in SDUs are displayed for each row on the right-hand side. In all cases, north is up and east to the left. The observational conditions of each image is given in Table 1. Images 1, 7, 10, 11, 18, 21, 24, 25, 28, and 30 are marked with †, which are used for comparison with the model in Fig. 8.

-2; see, for example, Rotundi et al. (2015), Fulle et al. (2016b), and Ott et al. (2017). The terminal velocities of the particles are parametrised by:

$$v(t, \beta) = v_0(t) \times \beta^\gamma, \tag{3}$$

where $\gamma = 1/2$. This assumption is commonly accepted for hydrodynamical drag from sublimating ice; see, for example, Moreno et al. (2011) and Licandro et al. (2013), which also agrees with the value range of 0.42 to 1.5 reported from GIADA measurements (Della Corte et al. 2015, 2016). On the other hand, $v_0(t)$ is a time-dependent term which is determined during the modelling process.

In addition to $v_0(t)$, the other time-dependent parameters in the model are the dust mass loss rate, $Q_{dust}(t)$, the maximum size of the particles, $r_{max}(t)$, and the power-law index of the particle size distribution $\delta(t)$. Due to the large number of free parameters used in the model, the obtained solution is not always unique, and it may be possible to find an alternative set of parameter values that also fit the observational data. However, this indetermination is reduced considerably when the available observations cover a significant orbital arc. The modelling method

Table 1: Observation Log.

| Date 2017 (UT)[1] | $r_h$ [2] (au) | $\Delta$ [3] (au) | Days to perihelion | Resolution[4] (km pixel$^{-1}$) | Dimension [5] (pixels$^2$) | Phase Angle (°) | $Af\rho$ [6] (cm) | Telescope[7] |
|---|---|---|---|---|---|---|---|---|
| (1†) January 20, 04:20 | -1.493 | 0.536 | -82.5 | 995.5 | 40 | 14.6 | 5.6±1.1 | TS |
| (2) January 30, 03:53 | -1.410 | 0.432 | -72.6 | 802.1 | 40 | 8.4 | 10.4±2.1 | TS |
| (3) February 5, 04:16 | -1.363 | 0.378 | -66.5 | 701.8 | 60 | 3.9 | 13.8±2.7 | TS |
| (4) February 16, 21:04 | -1.274 | 0.289 | -54.8 | 536.6 | 100 | 6.8 | 20.8±4.1 | TN |
| (5) February 19, 02:00 | -1.258 | 0.274 | -52.6 | 508.7 | 125 | 9.3 | 22.0±4.4 | TN |
| (6) February 25, 00:15 | -1.218 | 0.239 | -46.7 | 443.9 | 125 | 15.8 | 24.8±4.9 | TN |
| (7†) February 25, 04:22 | -1.217 | 0.238 | -46.5 | 441.9 | 125 | 16.0 | 25.8±5.1 | TS |
| (8) March 01, 21:42 | -1.186 | 0.214 | -41.8 | 397.3 | 150 | 21.9 | 21.1±4.2 | TN |
| (9) March 11, 20:00 | -1.130 | 0.175 | -31.9 | 324.4 | 100 | 35.5 | 18.6±3.7 | TN |
| (10†) March 14, 20:03 | -1.116 | 0.165 | -28.9 | 306.3 | 150 | 39.7 | 25.8±5.1 | TN |
| (11†) March 26, 01:03 | -1.072 | 0.144 | -17.7 | 267.4 | 200 | 55.4 | 27.5±5.5 | TN |
| (12) March 26, 22:46 | -1.069 | 0.144 | -16.8 | 267.4 | 200 | 56.5 | 29.6±5.9 | TN |
| (13) March 28, 02:19 | -1.066 | 0.143 | -15.6 | 265.5 | 175 | 56.7 | 26.4±5.2 | TN |
| (14) March 29, 23:52 | -1.061 | 0.142 | -13.7 | 263.6 | 150 | 61.2 | 26.2±5.2 | TN |
| (15) March 30, 22:31 | -1.059 | 0.142 | -12.8 | 263.6 | 150 | 61.2 | 26.3±5.2 | TN |
| (16) April 01, 01:37 | -1.057 | 0.142 | -11.6 | 263.6 | 175 | 62.3 | 24.5±4.9 | TN |
| (17) April 03, 21:01 | -1.052 | 0.142 | -8.8 | 263.6 | 150 | 64.9 | 23.6±4.7 | TN |
| (18†) April 07, 23:12 | -1.047 | 0.145 | -4.8 | 269.2 | 150 | 67.8 | 23.0±5.0 | TN |
| (19) April 12, 23:52 | 1.045 | 0.152 | 0.2 | 282.2 | 150 | 69.7 | 23.6±5.2 | TN |
| (20) April 19, 22:56 | 1.049 | 0.165 | 7.5 | 306.3 | 150 | 69.7 | 25.3±5.0 | TN |
| (21†) April 21, 23:41 | 1.052 | 0.169 | 9.2 | 313.8 | 150 | 69.3 | 26.3±5.2 | TN |
| (22) April 26, 04:49 | 1.061 | 0.179 | 13.4 | 332.3 | 150 | 67.6 | 28.3±5.6 | TN |
| (23) April 27, 22:58 | 1.065 | 0.184 | 15.2 | 341.6 | 150 | 66.6 | 28.7±5.7 | TN |
| (24†) May 02, 23:59 | 1.080 | 0.197 | 20.2 | 365.7 | 130 | 63.4 | 27.3±5.4 | TN |
| (25†) June 07, 03:12 | 1.278 | 0.319 | 55.3 | 592.3 | 60 | 30.3 | 23.0±4.6 | TN |
| (26) June 11, 04:34 | 1.308 | 0.339 | 59.4 | 629.4 | 60 | 26.5 | 21.6±4.3 | TN |
| (27) June 22,00:30 | 1.392 | 0.403 | 70.2 | 374.1 | 100 | 18.1 | 24.7±4.9 | TN |
| (28†) June 29, 02:50 | 1.450 | 0.455 | 77.3 | 422.4 | 100 | 14.9 | 19.7±3.9 | TN |
| (29) July 20, 20:00 | 1.633 | 0.665 | 99.0 | 617.3 | 50 | 17.0 | 10.5±2.1 | TN |
| (30†) July 27, 23:27 | 1.695 | 0.752 | 106.2 | 698.1 | 50 | 19.2 | 8.9±1.7 | TN |

**Notes.**
(1) Dates marked with a † are used for comparison with the model in Fig. 8.
(2) Heliocentric distance. Negative values correspond to pre-perihelion, positive values to post-perihelion.
(3) Geocentric distance.
(4) Resolution of the images in Fig. 1.
(5) Dimensions of the images in Fig. 1.
(6) $\rho = 10^4 km$
(7) TS corresponds to TRAPPIST-South, TN to TRAPPIST-North.

consists of a trial-and-error procedure starting from the simplest scenario, i.e., symmetric behaviour of the time-dependent parameters with respect to perihelion, assuming an isotropic ejection pattern. From this starting point, we subsequently varied the parameter values. If after many trials a good match with the observations is not found, we switch to anisotropic ejection pattern, where the emission of the particles is characterised by active areas on the comet's surface and the rotational state is defined by by two angles (Sekanina 1981): the obliquity of the orbit plane to the equator, $I$, and the argument of the subsolar meridian at perihelion, $\phi$. The obliquity determines the sense of the rotation, which is prograde when $0° \leq I < 90°$ and retrograde when $90° < I \leq 180°$. On the other hand, when $0° < \phi < 180°$, the northern pole experiences sunlight at perihelion; the southern pole when $180° < \phi < 360°$. However, there is an ambiguity in the formalism, and equivalent solutions are given by $180° - I$ and $\phi \pm 180°$, thus the sense of the rotation remains indeterminate.

To determine which is the best model, we followed the method of Moreno et al. (2016a), who computed the goodness-of-fit through the quantity $\chi = \sum \sigma_i$, where the summation extends to all images under consideration, and:

$$\sigma_i = \sqrt{\sum [log(I_{obs}(i)) - log(I_{mod}(i))]^2 / N(i)}, \quad (4)$$

where $N(i)$ is the number of pixels in the image $i$, and $I_{obs}(i)$ and $I_{mod}(i)$ are the observed and modelled brightnesses, respectively. For every trial, we calculated $\sigma$ for each image looking for the minimum $\chi$. The choice of work in the logarithm of the intensities instead of the original intensities, allow us to give more appropriate weights to the outermost and innermost isophotes.

Due to a lack of information about the nucleus itself, we had to assume its bulk density, $\rho_N$, and size, $R_N$. Pätzold et al. (2016) estimated that the average bulk density of the nucleus of 67P was $\rho = 533 \pm 6$ kg m$^{-3}$, based on Rosetta measurements. From the Deep Impact mission, A'Hearn et al. (2005) estimated the nucleus density of comet Tempel 1 to be $\rho_N \sim 600$ kg m$^{-3}$. Since comet 41P belongs to the same family of comets, some properties are expected to be roughly similar; therefore, we assumed an intermediate value of $\rho_N \sim 550$ kg m$^{-3}$. On the other hand, we find that the estimated the size of 41P's nucleus to be $R_N \sim 0.7$ km, and Howell et al. (2017) reported a minimum size of $R_N \sim 0.9$ km from radar observations. We adopted the latter value in this study.

## 3.2. Results and discussion

We found that we could not explain the observational data set using a full isotropic ejection model, and it was necessary to switch to a more complex scenario. In fact, we find that the emission pattern experienced two transformations: from full isotropic to anisotropic dominated by two strongly active areas, and then from anisotropic to full isotropic again. We named this the hybrid model. A comparison between full isotropic emission and the hybrid model is shown in Fig 2.

Chronologically, the comet started with a full isotropic ejection pattern, when the level of activity was very low and it had dust production rates of $Q_{dust} = (3 - 20)$ kg s$^{-1}$. When the comet was ∼(1.218-1.116) au inbound (February 24-March 14, 2017), the emission started to switch to anisotropic. Two active areas took over and dominated the emission of the dust particles, ejecting ∼ 90% of them. The latitudes of these active areas were found to be located both in the northern hemisphere (from $(45 \pm 10)°$ to $(90^{+0}_{-10})°$) and in the southern hemisphere (from $(-35 \pm 10)°$ to $(-90^{+10}_{-0})°$). The combination of these areas represents about 23.7% and 51.2% of the total cometary nucleus surface, respectively. In principle, from our models, we are also able to constrain the longitude of the active areas. However, a well defined area in both longitude and latitude on the surface of a rotating nucleus, after many rotations, will produce the same effect as if the nucleus was ejecting particles across the entire longitude range (from 0° to 360°). Therefore, we were only able to properly constrain the latitudes of the active areas, and we are missing information about their longitudes.

This ejection pattern lasted until ∼(1.20-1.45) au outbound (June 7-28, 2017), when the activity decreased and switched again to full isotropic. During the anisotropic period, the level of activity reached a maximum a few days before perihelion, and it was possible to characterise the rotational state of the comet via rotational angles $\phi$ and $I$. The subsolar meridian at perihelion was found to be $\phi = (5 \pm 3)°$, and the obliquity $I = (25 \pm 15)°$. A schematic vision of the hybrid model is displayed in Fig. 3.

From aperture photometry performed on CN narrowband imaging, Bodewits et al. (2018) found a very rapid change in the apparent periodicity of the jets from 20 to 50 hours between March and May 2017. In Moulane et al. (2018) we confirm this behaviour through TRAPPIST imaging by the comparison of coma features exhibited by the CN gas species between March and April 2017. Rotational period changes are not uncommon in comets, however, it is expected that such changes take place over a time period of a number of years, being observable from one orbit to another (Samarasinha et al. 2004). Bodewits et al. (2018) found two jets associated with 41P, which were proposed as the main cause of the spin down.

In this context, our dust model with two strongly active areas seems to match well with their results. In order to explain the spin down, these powerful dust jets, which imply strong outgassing, should be located at different longitudes in the north and in the south, in such a way that when combined with the small size of the nucleus, they exert torques that act as brakes due to reactional forces. However, as explained before, we are unable to properly constrain the longitude of the active areas and hence unambiguously confirm this assertion.

On the other hand, in order to provoke a spin down in a cometary nucleus, the most efficient place would be at the equator. Unfortunately, all of our attempts to place an active area at the equator were unsuccessful. In addition, the local topography and/or the real physical shape of the cometary nucleus could also greatly affect the final net torque. However, the lack of information about the nucleus itself prevents us from confirming this hypothesis, and leads us to consider that other unknown factors could also be affecting the final torque. Moreover, since the rotational periods found by Bodewits et al. (2018) are much shorter than the age of the dust tail, it is not possible to confirm these values with our dust model.

The evolution of the dust parameters that best describe the dust environment along its orbit are displayed in Fig. 4, the dust production rate; Fig. 6, power index of the size distribution and maximum size of particles; and Fig. 7, ejection velocities of the particles for several sizes.

Overall, we found that the activity started about 180±10 days before perihelion (October 14, 2016), which corresponds to 2.31±0.08 au inbound, that is, between October 4 and October 24, 2016. The activity increased gently until -1.64 au, whereby it began to increase to a faster rate until the peak of the activity at ∼ 17 days pre-perihelion with a dust production rate of $Q_{dust} = 110$ kg s$^{-1}$. The total dust ejected from the estimated activation date to our last observation is ∼ $7.5 \times 10^8$ kg. Since our observations cover a range that spans the beginning of the activity until when the comet nearly switched off, we can consider that this is a good estimate of the total dust ejected during the complete orbit. This value is significantly lower than previously reported for other comets of the same family also considering a complete orbit; see, for example, 67P, $1.4 \times 10^{10}$ kg given by Moreno et al. (2017a), 81P/Wild 2 (hereafter 81P), $1.1 \times 10^{10}$ kg given by Pozuelos et al. (2014a) or 22P/Kopff, with a total of $8 \times 10^9$ kg estimated by Moreno et al. (2012). Comet 41P is therefore a dust-poor comet (Fig 4). However, one has to keep in mind that 41P is remarkably smaller in size than these comets, where its total mass is estimated to be M∼ $1.6 \times 10^{12}$ kg from the current data available and the hypotheses made in this study regarding its nucleus density and spherical shape. This means a small but non-negligible amount of erosion occurs per perihelion passage. The dust-to-gas production rate ratio provides important clues about the formation mechanism and evolution of comets. Using the TRAPPIST telescopes, we also carried out observations of the main gases: OH, CN, NH, $C_2$, and $C_3$. Those observations and the gases' production rates will be presented in a separate paper (Moulane et al. 2018). In particular, from the production rates of OH, we computed the water production rates using the relation:

$$Q_{H_2O} = 1.361 \times r^{-0.5} \times Q_{OH}, \qquad (5)$$

given by Cochran & Schleicher (1993). In Fig. 4, both the dust production and water production rates are displayed, while in Fig. 5 the evolution of the dust-to-water mass ratio is shown, which varies from 0.5 at ∼(1.3-1.10) au inbound to 1.5 when the peak of activity reached a maximum a few days before perihelion at 1.07 au. Post-perihelion, the dust-to-water mass ratio stayed at values close to 1 until the last observation available, when it dropped to 0.25. This low dust-to-water mass ratio implies, and confirms, the aforementioned assertion that 41P is a dust-poor comet, and its nucleus is richer in volatiles. These values are notably smaller than those obtained for 67P, where the dust-to-water mass ratios ranged from 6 to 100 at perihelion and immediately afterwards from *in situ* measurements obtained with the GIADA and OSIRIS instruments inferred by individual particle detections (Fulle et al. 2016b), and 4±2 in Rotundi et al. (2015). Comets with low dust-to-gas mass ratios are more prone to outbursts, which is attributed to a mechanism based on the phase transitions of volatile species between solid, liquid, and gaseous states. Moreover, high mixing ratios of volatile ices may indicate low physical strengths, where dust fails to act as an efficient binder, thus provoking an unstable nucleus prone to fragmentation events (Miles 2016). An example of this is comet 73P/Schwassmann-Wachmann, which split into many fragments in 2006 (Ishiguro et al. 2009); its dust-to-gas ratio was 0.12 (Lisse et al. 2002). 41P has suffered several outbursts in its recent past: two violent outbursts in 1973, with a brightness amplitude of 9 magnitudes (Kresak 1974) and two other outbursts in 2001, with a brightness amplitude of 4 magnitudes (Bodewits et al. 2018). Due to the dust erosion per orbit found in our model for such a small nucleus, the low dust-to-gas ratio and its propensity to outbursts, 41P is a good candidate for a split event in the future.

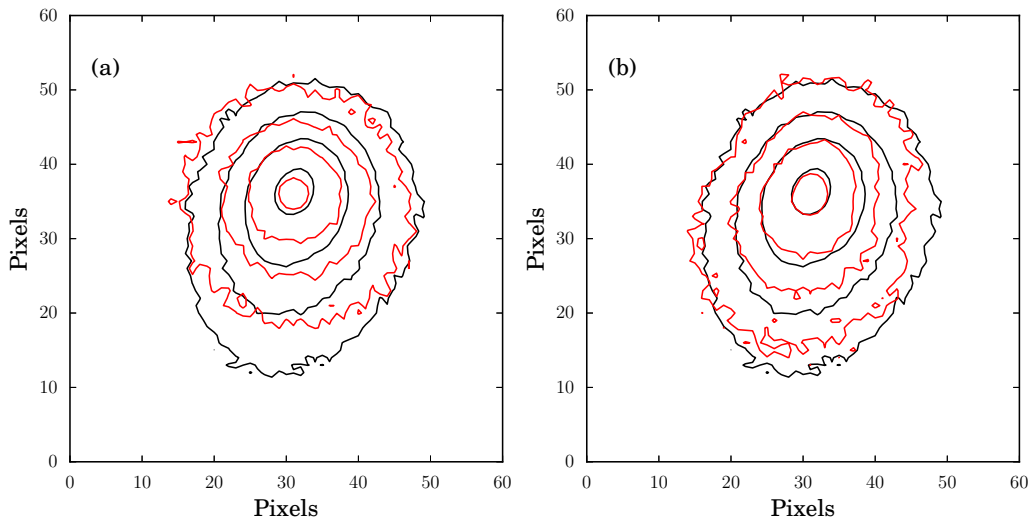

Fig. 2: Comparison between the full isotropic model, panel (a), and the hybrid model, panel (b). In both cases, the red contours correspond to the models and the black ones to the observations. The observation date is March 14, 2017. For modelling purposes, the image was rebinned ×2 with respect to the value given in Table 1. Therefore, the resolution is 612.6 km pixel$^{-1}$. In all cases the isophote levels are: $4×10^{-14}$, $7×10^{-14}$, $1.25×10^{-13}$ and $4×10^{-13}$ SDU. The validity of the models, determined via the $\chi$ parameter, confirms that the isotropic model offers a poorer fit ($\chi_{isotropic} = 3.9$) than the hybrid model ($\chi_{hybrid} = 2.4$). The plot is orientated so that north is up and east is to the left.

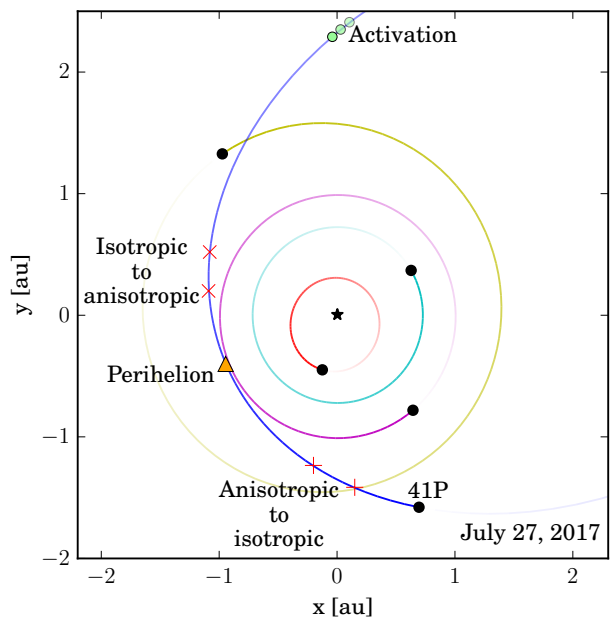

Fig. 3: Schematic vision of the evolution of the dust emission pattern of 41P along its orbit. The four inner planets in the solar system are also included. The configuration corresponds to the time of our last observation, that is, July 27, 2017. The two pairs of red crosses point out the changes in the emission pattern: first is from full isotropic to anisotropic during the inbound journey (February 24-March 14, 2017) and the second is from anisotropic to full isotropic during the outbound journey (June 7-28, 2017). The estimated start date of the activity (October 14, 2016) is marked with green dots, and perihelion (April 12, 2017) is marked with an orange triangle.

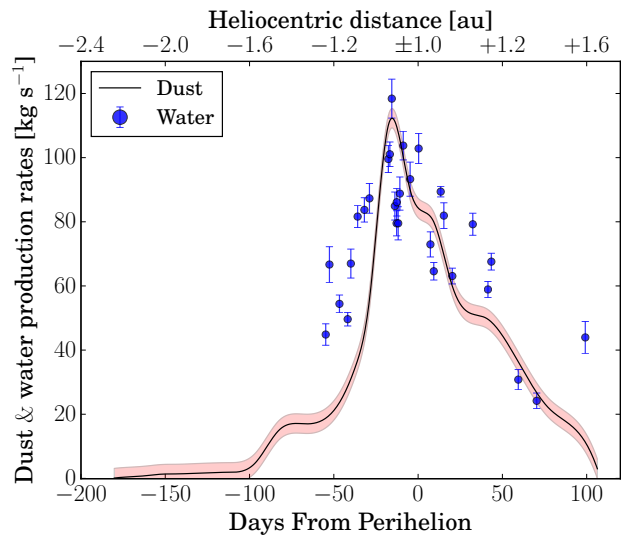

Fig. 4: Dust production rate (solid black line) and its error (red border) given by the hybrid model as a function of day relative to perihelion (lower $x$-axis) and heliocentric distance (upper $x$-axis). The water production rates computed from OH observations with TRAPPIST telescopes are shown as blue dots (Moulane et al. 2018).

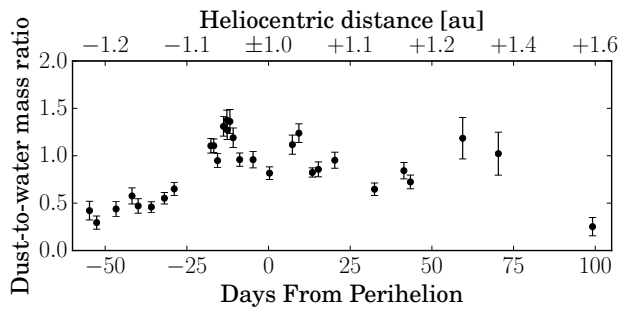

Fig. 5: Dust-to-water mass ratio as a function of day relative to perihelion (lower x-axis) and heliocentric distance (upper x-axis).

The power-law index of the size distribution at the beginning of the activity was $\delta = -3.50$. It reached a minimum around perihelion as $\delta = -3.75$ and finished as $\delta = -3.60$ (Fig. 6, panel (a)). This behaviour of having a minimum value at perihelion was reported previously in other comets; see, for example, 67P in Moreno et al. (2017a), and 81P and 103P/Hartley 2 (103P) in Pozuelos et al. (2014a), which indicates that the population of small particles ejected increase and dominate the size distribution and the observed brightness. The minimum size of particles was a constant in the model presented here (set to 10 $\mu$m), and it is these particles that were the fastest, reaching $v_{ejec} \sim 200$ m s$^{-1}$ at perihelion. On the other hand, the slowest velocity in the model corresponds to the largest particles, which increase from 1 cm at the start of the activity to 40 cm during the peak of the activity, and decrease to 2 cm at the end. We called this velocity $V_{min}$ (Fig. 6, panel (b) and Fig. 7). The maximum size of the particles found during perihelion agrees very well with the results from *in situ* missions such as Rosetta, where the authors found particles up to 40 cm at perihelion distances (Rotundi et al. 2015). Values of 20 cm were reported by Harmon et al. (2011) during the fly-by of *EPOXI* mission (A'Hearn et al. 2011; Meech et al. 2011) around 103P. The minimum velocity in the model, $V_{min}$, has to fulfil the condition $V_{min} \gtrsim V_{esc}$, where $V_{esc}$ is the escape velocity of the nucleus given by

$$V_{esc} = R_N \sqrt{\frac{2}{15}\pi\rho G},$$ (6)

for spherical shape, at distance $\sim 20$ $R_N$ where the gas drag vanishes. We find $V_{esc} \sim 0.01$ m s$^{-1}$, the condition being overcome all the time. The velocities reported by Della Corte et al. (2016) from GIADA measurements, adopted by Moreno et al. (2017a) (their Fig. 2), showed particles of 100 $\mu$m at $\sim 1.5$ au with velocities of $\sim 22$ m s$^{-1}$. In this study we derive that, at the same heliocentric distance and for the same size of particles as 67P, 41P ejected particles at about $\sim 10$ m s$^{-1}$ (see Fig. 7). This is consistent with the fact that 41P is smaller in size, and it has both weaker dust production and gas production rates than 67P. This can be concluded by comparing the values obtained for 67P by Opitom et al. (2017) and the values obtained for 41P by Moulane et al. (2018). Both studies used TRAPPIST telescopes.

A comparison of ten selected images from the observational data set (panels marked with a † in Fig. 1) with the corresponding synthetic images generated by the model are shown in Fig. 8. In addition to the images, our model was forced to match the observationally derived $A(\theta)f\rho$ parameter (A'Hearn et al. 1984). This quantity is related to the dust coma brightness, where $A(\theta)$ is the geometric albedo as a function of phase angle ($\theta$), $f$ is the filling factor in the aperture of field of view, and $\rho$ is the projected distance from the nucleus. We computed $A(\theta)f\rho$ for our complete set of observations at $\rho = 10^4$ km, which allowed us to track its time evolution.

Since the phase angle varied significantly from 0° to 70° (see Table 2), it was necessary to adopt a correction to distinguish between enhancements due to the actual behaviour of the comet and those arising from phase angle effects, such as backscattering at low phase angles. This effect has been observed for a number of comets. In particular, for 67P it was observed through ground-based observations made by (Moreno et al. 2017a), where the authors reported two enhancements at low phase angles. The first of them occurred 400 days pre-perihelion, when it was almost null because of the low level of activity of the comet at that moment. The second peak occurred 200 days post-perihelion, when the level of activity was higher and the enhancement was better resolved. Moreover, this effect was confirmed by Bertini et al. (2017), where the authors studied the scattering phase function at a wide range of phase angles both pre- and post-perihelion *in situ* using the OSIRIS instrument on board Rosetta. However, in our $A(\theta)f\rho$ measurements the backscattering effect does not seem appreciable. There may be several reasons for this behaviour: it may be due to the very low dust production rate at the time of the minimum, which occurred at the beginning of the activity, and therefore the total amount of dust was very low and the precision of the data are not good enough. It may also be due to the low number of observations available around the minimum. These reasons suggest that the expected backscattering enhancement was not detectable in our dataset. Moreover, it may have been hidden because of the rapid increase of cometary activity a short while afterwards. Notwithstanding, we adopted a correction following the combined phase function computed by D. Schleicher[1] from observations of comets at all phase angles.

The values of $A(\theta)f\rho$ directly computed from the observations, and the conversion to $A(\theta = 0°)f\rho$, are compared with the synthetic values obtained from the hybrid model and displayed in Fig. 9. In general terms, the hybrid model proposed here fits very well both images and the $A(\theta = 0°)f\rho$ parameter.

## 4. Dynamical analysis

The main reason for studying the dynamical evolution of comet 41P is to understand how long this comet has been suffering its current rate of erosion, and for how long it will continue. Comet 41P is a near-Earth Jupiter family comet (NEJFC), that is, a JFC with a perihelion distance of $q < 1.3$ au. The recent work by

---

[1] http://asteroid.lowell.edu/comet/dustphase_details.html

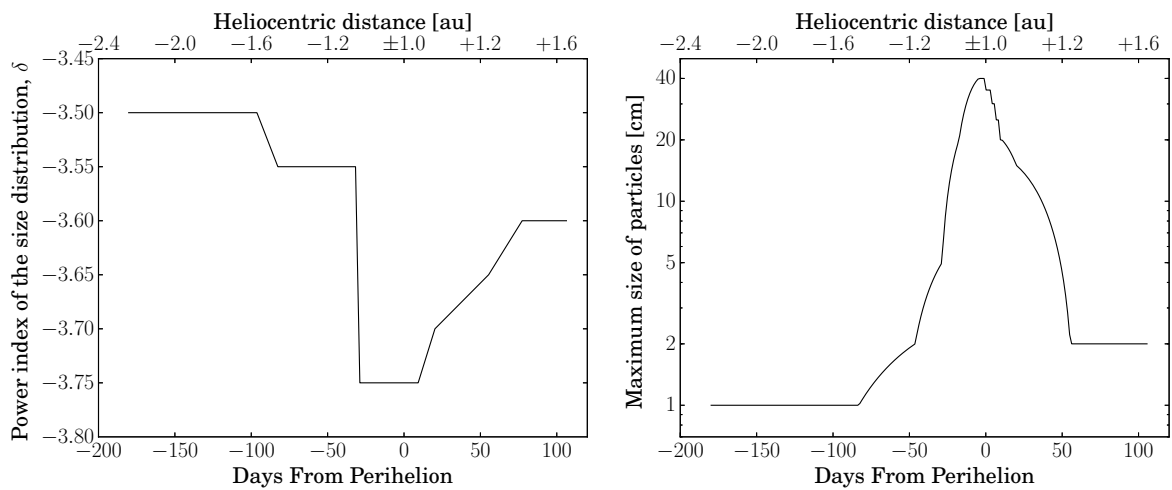

Fig. 6: Left panel: evolution of the power-law index of the particle-size distribution, $\delta(t)$. Right panel: evolution of the maximum size of the ejected particles, $r_{max}(t)$. Both are presented as functions of day relative to perihelion (lower $x$-axis) and heliocentric distance (upper $x$-axis).

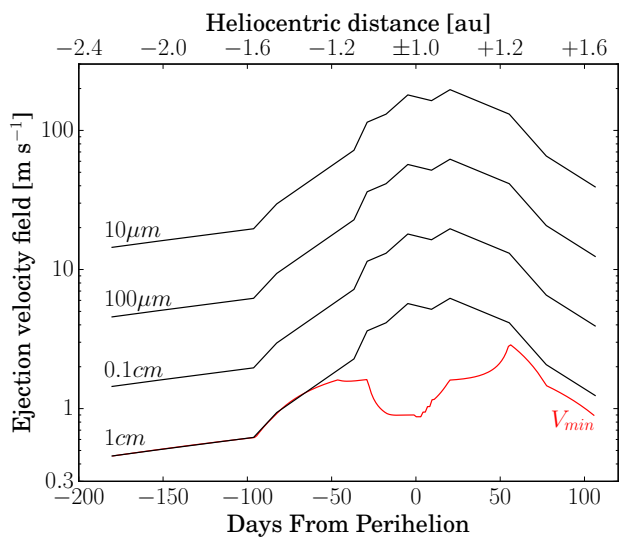

Fig. 7: Ejection-velocity field of the hybrid model, as a function of day relative to perihelion (lower $x$-axis) and heliocentric distance (upper $x$-axis). The velocity is parametrised in the model by equation (3). The fastest particles correspond to a size of 10 $\mu$m. In addition, the velocities of the 100 $\mu$m, 0.1 cm, and 1 cm sized particles are shown. The slowest velocity in the model is red-labelled as $V_{min}$, which corresponds to the largest particles (from 1 cm to 40 cm; see Fig. 6, right panel).

Fernández & Sosa (2015) revealed a subgroup among NEJFCs that reside in highly stable orbits, with a likely origin in the main asteroid belt. This new class of objects could be the counterparts to the Main Belt Comets (Jewitt et al. 2015), that is, they may be asteroids disguised as comets. In order to clarify the dynamical nature of 41P, we performed numerical integrations like many other authors before; see, for example, Di Sisto et al. (2009), Ye et al. (2016), and Fernández et al. (2017).

## 4.1. Numerical integrations

In order to study the dynamical evolution of 41P, we used numerical integrations in the heliocentric frame. A quick first inspection consisted of integrations over $2 \times 10^4$ yr: from current time to $10^4$ yr backward and $10^4$ yr forward. The current time was set as October 25, 2017. The initial conditions of the orbital parameters were extracted from the NASA/JPL Small-Body Database, and can be consulted in Table 2. The numerical integrations were performed twice using different numerical packages, MERCURY6 (Chambers 1999) and REBOUND (Rein & Liu 2012; Rein & Spiegel 2015) with equivalent conditions. The results obtained were the same; therefore here we only describe the ones obtained with MERCURY6, which allow for a more meaningful comparison with analyses performed by other authors. The Sun and the eight planets were included in the simulations. We used the hybrid algorithm, which combines a second-order mixed-variable symplectic algorithm with a Burlisch-Stoer integrator to manage the close encounters. The initial time-step was set to 8 days, and the computed orbital evolution was stored every year. We considered as ejected the particles with heliocentric distance $r_h > 100$ au. Any close encounters with planets at distances smaller than 3× Hill radii were also registered. From the dust analysis, we derived the action of two strongly active areas which seem to be related to the fast rotational period variations. We decided to perform two trials: the first one considered only a pure gravitational model, and the second included non-gravitational forces (see table 2). In the first experiment, the results were exactly the same, therefore only the results for the pure gravitational model are shown in Fig. 10. From these results, we find that its current state as an NEJFC was obtained in the recent past after a close encounter with Jupiter in the period of time studied ($\sim -1572$ yr), at a distance of $d_{41P-Jupiter} \sim 0.12$ au, which is well below the Hill radius. Due to this event, all the orbital parameters changed dramatically. In addition, it was observed that currently the comet is in a transitional state, and it may reach a more stable orbit with some of its orbital parameters coupled. The perihelion distance would be even smaller in that future scenario, with a value of $q \sim 0.8$ au. Therefore, we conclude that being a NEJFC seems to be a relatively new status for 41P, and it will remain as part of this population of comets for at least the next $10^4$ yr.

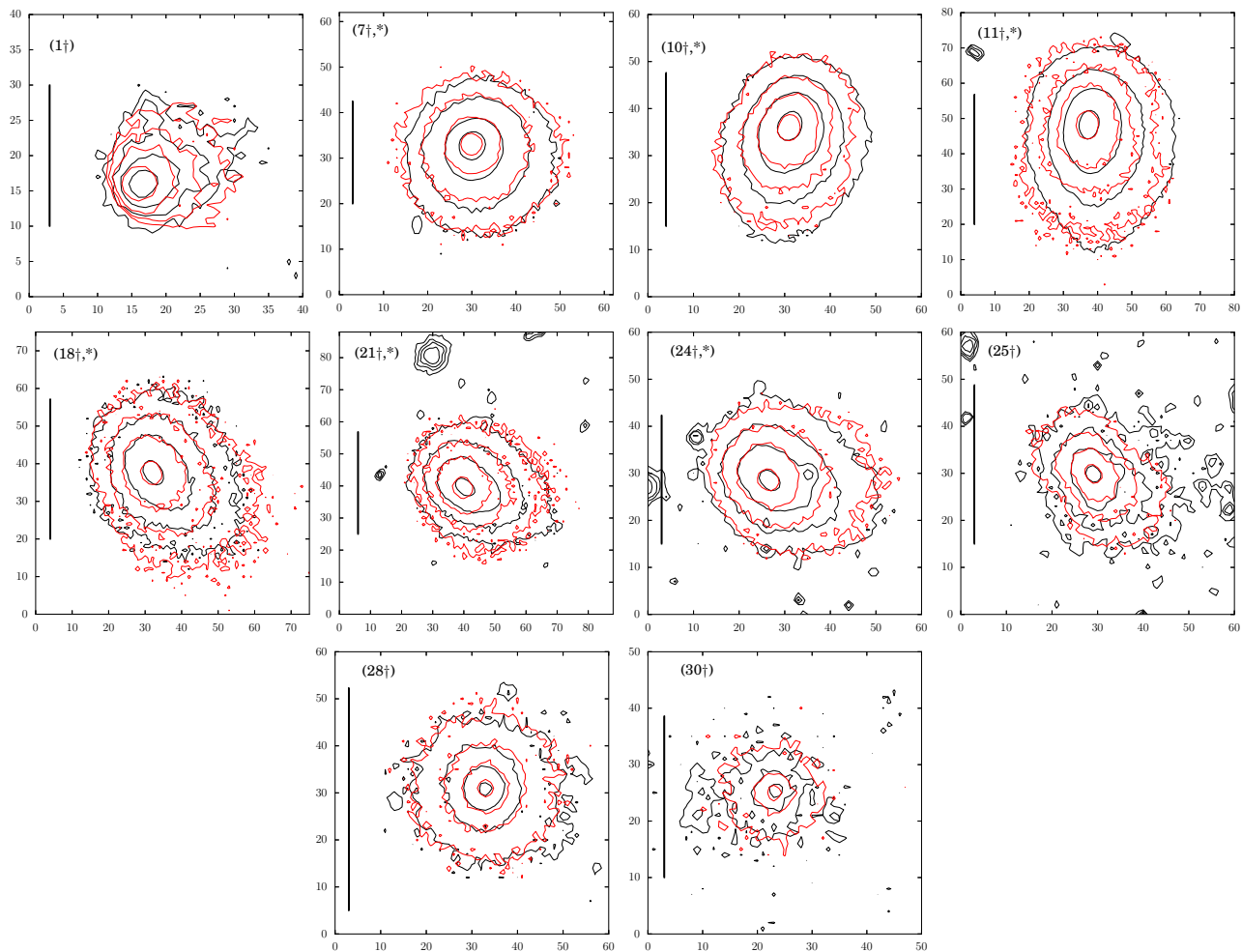

Fig. 8: Comparison of a subset of ten observed and modelled images. The selected observed images are marked with a † in Fig. 1; here we keep the same numeration. For modelling purposes, those images labelled with an '\*' were rebinned ×2 with respect to the values given in Table 1. The isophote levels in SDU in each case are: (1†) $0.5 \times 10^{-14}$, $1.0 \times 10^{-14}$, $2.0 \times 10^{-14}$ and $6.0 \times 10^{-14}$. (7†,\*) $1.7 \times 10^{-14}$, $3.0 \times 10^{-14}$, $7.0 \times 10^{-14}$ and $2.0 \times 10^{-13}$. (10†,\*) $4.0 \times 10^{-14}$, $7.0 \times 10^{-14}$, $1.2 \times 10^{-13}$ and $4.0 \times 10^{-13}$. (11†,\*) $4.0 \times 10^{-14}$, $7.0 \times 10^{-14}$, $1.2 \times 10^{-14}$ and $4.0 \times 10^{-13}$. (18†,\*) $4.0 \times 10^{-14}$, $7.0 \times 10^{-14}$, $1.2 \times 10^{-14}$ and $4.0 \times 10^{-13}$. (21†,\*) $4.0 \times 10^{-14}$, $7.0 \times 10^{-14}$, $1.2 \times 10^{-14}$ and $4.0 \times 10^{-13}$. (24†,\*) $4.0 \times 10^{-14}$, $7.0 \times 10^{-14}$, $1.2 \times 10^{-14}$ and $4.0 \times 10^{-13}$. (25†) $4.0 \times 10^{-14}$, $7.0 \times 10^{-14}$, $1.2 \times 10^{-14}$ and $4.0 \times 10^{-13}$. (30†) $1.7 \times 10^{-14}$, $3.5 \times 10^{-14}$ and $9.0 \times 10^{-14}$. In all cases, the black contours correspond to observations and the red ones to the model. The black vertical lines correspond to $2 \times 10^4$ km in the sky plane, and the $y$ and $x$ axes are given in pixels.

However, the chaotic nature of the orbits of the minor bodies in the solar system is well known (Levison & Duncan 1994). Therefore, in order to make more robust conclusions, it is necessary to perform statistical studies. With this aim, we carried out the set of tests proposed by Fernández et al. (2014) and Fernández & Sosa (2015) to obtain the instability degree of comet 41P.

### 4.2. Characterisation of the degree of instability of 41P's orbit

While typical JFCs have unstable orbits, probably coming from trans-Neptunian regions, a small group of them reside in asteroidal orbits, and may even originate from the asteroid belt, like near-Earth asteroids. Fernández & Sosa (2015) hypothesised that some NEJFCs could be interlopers from the asteroid belt. These authors analysed a sample of 58 NEJFCs using numerical simulations of their orbits in order to characterise their stability, and they performed statistical studies using clones of those NEJFCs. They found different degrees of stability, which allowed them to distinguish three different categories: *Highly asteroidal*, whose orbits are highly stable with a possible source region being the outer main asteroid belt; *Moderately asteroidal*, despite generally having highly stable orbits, a small fraction of the clones were unstable, and therefore gave less confidence to their origin; and *Maybe asteroidal*, predominantly displaying stable orbits, but an important fraction of the clones had highly unstable orbits due to the occurrence of encounters with Jupiter at distances of less than 0.2 au, more in consonance with the evolution of JFCs. Therefore, a possible asteroidal origin is more uncertain.

Fernández & Sosa (2015) included 41P in their sample. However, they did not report any peculiar result regarding it; therefore, we infer that they did not find anything suspicious to suggest that this comet originated in the asteroid belt. However, in the last perihelion passage in April 2017, the extremely

Table 2: Orbital parameters of comet 41P/Tuttle-Giacobini-Kresak.

| | |
|---|---|
| Epoch | 2457844.5 (2017-April-01.0) TDB |
| Perihelion date | 2457856.25 (2017-April-12.75) |
| Orbital Period | 5.42 [yr] |
| Perihelion distance ($q$) | 1.04504273±3.3×10⁻⁸ [au] |
| Semimajor axis ($a$) | 3.085003±3.4×10⁻⁶ [au] |
| Eccentricity ($e$) | 0.6612506±3.6×10⁻⁷ |
| Inclination ($i$) | 9.229131±2.6×10⁻⁶ [deg] |
| Argument of perihelion ($\omega$) | 62.15858±1.9×10⁻⁵ [deg] |
| Longitude of ascending node ($\Omega$) | 141.06628±1.4×10⁻⁵ [deg] |
| Non-gravitational radial acceleration ($A1$) | 1.7452037294 × 10⁻⁸±2.1×10⁻¹⁰ [au day⁻²] |
| Non-gravitational transverse acceleration ($A2$) | 4.2753305920 × 10⁻⁹±2.1×10⁻¹⁰ [au day⁻²] |
| Non-gravitational normal acceleration ($A3$) | 1.4734309689 × 10⁻⁹±1.1×10⁻¹⁰ [au day⁻²] |

**Notes.**
Osculating values of orbital parameters ±1$\sigma$ uncertainty.
Source: JPL Small-Body Database (JPL K171/18).

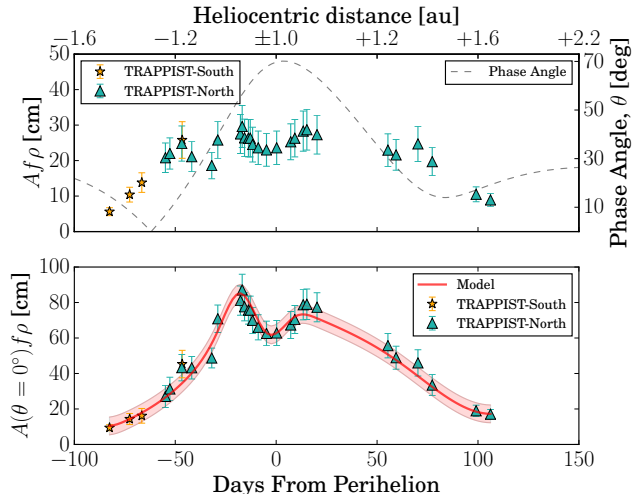

**Fig. 9:** Upper panel: $A(\theta)f\rho$ values computed directly from the observational data set (left $y$-axis) and phase angle, $\theta$ (right $y$-axis). Lower panel: $A(\theta = 0°)f\rho$ observational values and the synthetic $A(\theta = 0°)f\rho$ values obtained from the hybrid model. Both panels show the data as a function of day relative to perihelion (lower $x$-axis) and heliocentric distance (upper $x$-axis).

favourable conditions for its observation attracted international attention, increasing the number of observations available, and improving the quality of its determined orbital parameters. Its current condition code is 2 [2]. Therefore, we decided to re-analyse its orbital stability following the same steps given by Fernández & Sosa (2015), where a *likely dynamical path* is defined as the average of the set of results obtained for a given object and its clones, characterised by the $f_q$ index, $f_a$ index, the capture time (here after $t_{cap}$), and the closest approach to Jupiter, $d_{min}$. Here, we briefly describe these parameters, and we refer the reader to Fernández & Sosa (2015) and Fernández et al. (2014), and references therein, for further information.

The $f_q$ index is computed as:

$$f_q = \frac{\sum_{j=1}^{N+1} \Delta t_j}{(N+1) \times 10^4}, \tag{7}$$

where $\Delta t_j$ is the fraction of time in the last $10^4$ yr in which a given JFC or its clones ($j = 1, ..., N$) describe an orbit with $q > 2.5$ au or reaches heliocentric distances $r_h > 100$ au.

The $f_a$ index is computed as:

$$f_q = \frac{\sum_{j=1}^{N+1} \Delta t'_j}{(N+1) \times 10^4}, \tag{8}$$

where $\Delta t'_j$ is the fraction of time in the last $10^4$ yr in which a given JFC or its clones ($j = 1, ..., N$) move along an orbit with semi-major axis $a > 7.37$ au, i.e., an orbital period > 20 yr. When reaching these criteria, an object is no longer considered to be a JFC.

In general terms, comets in unstable orbits have $f_q$ and $f_a$ values well above zero, which means that they spend an important fraction of the $10^4$ yr with $q > 2.5$ and/or $a > 7.37$ au. On the other hand, when $f_q \sim f_a \sim 0$, the comets move in stable orbits.

The $t_{cap}$ parameter is defined as the time in the past at which the average perihelion distance of a given NEJFC and its clones at a certain time, that is,

$$\bar{q}(t) = \frac{\sum_{j=1}^{N+1} \Delta q_j(t)}{N+1}, \tag{9}$$

increased by 1 au with respect to the observed value at the discovery time, $t_{disc}$, namely:

$$q(t_1) = q(t_{disc}) + 1 \implies t_{cap} = t_{disc} - t_1. \tag{10}$$

An increase of 1 au in $q$ means that the comet is $q \sim 2$ au farther from the Earth's vicinity. Thus, the concept of $t_{cap}$ is related to the time span during which the comet has been in the Earth's region, i.e. making it a NEJFC. For instance, typical JFCs have $t_{cap}$ ranging from a few years to a few times $10^3$ years. On the other hand, the NEJFCs in the category defined by Fernández & Sosa (2015) as *Highly asteroidal* show values for $t_{cap}$ that largely exceed $10^4$ yr.

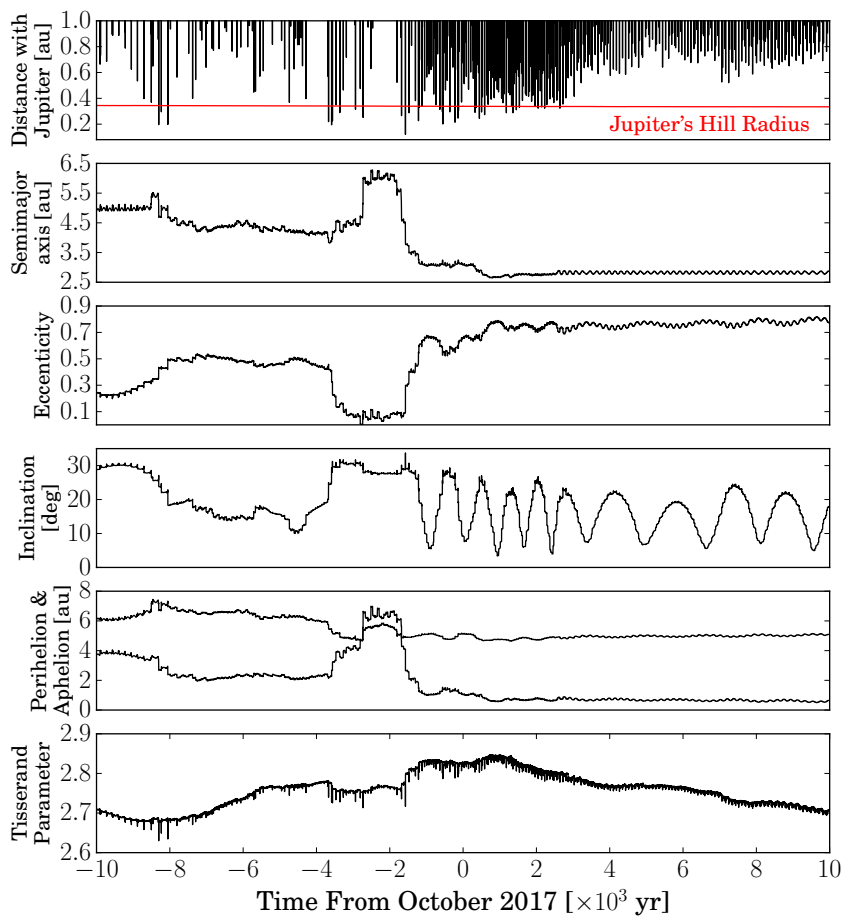

**Fig. 10.** Orbital evolution of 41P during $2\times10^4$ yr: from current time to $10^4$ yr backward and $10^4$ yr forward in time. From the top to the bottom: the closest approaches with Jupiter, semi-major axis, eccentricity, inclination, perihelion and aphelion distance, and Tisserand parameter. The initial orbital elements are displayed in Table 2, which were taken from the JPL Small-Body Data Browser.

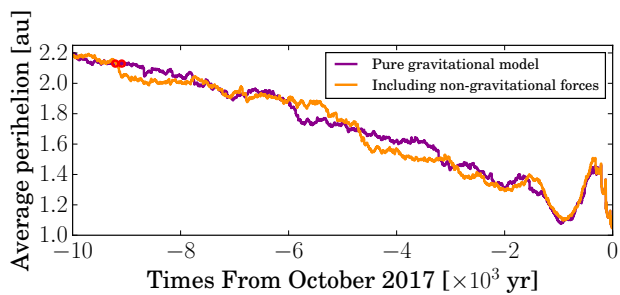

Fig. 11: Average perihelion distance as a function of time for both the pure gravitational model (magenta solid line) and the model that includes non-gravitational forces (yellow solid line). The red circles indicate the time when $\bar{q} = q_{disc} + 1$, which is necessary to compute the capture time (see text).

### 4.3. Results and discussion

In order to achieve a balance between computational cost and good coverage of the physical space around 41P, we generated 200 clones of the original 41P comet. The orbital parameters $a$, $e$, and $i$ were chosen randomly from a Gaussian distribution, whose mean values and standard deviations were their osculating values and their $3\times$ osculating uncertainties, $\sigma$, respectively (see Table 2). With this choice, we ensured that most of our clones ($\sim$ 70%) were in the $3\times\sigma$ area around the osculating orbital parameters. The total integration time was $10^5$ yr: from current time to $5\times10^4$ yr backward and $5\times10^4$ yr forward. We applied the concepts described above to our simulations with 200 clones for both models (pure gravitational and one that includes non-gravitational forces). We found very similar results for both of them. In the first instance, for a purely gravitational model, we obtained: $f_q$=0.025, $f_a$=0.007, and for that with non-gravitational forces, $f_q$=0.027, $f_a$=0.004. The $t_{cap}$ parameters were computed for both models from the average perihelion in the last $10^4$ yr given by Eq. (9), and they are displayed in Fig. 11. We find for the pure gravitational model $t_{cap}$=1.09×$10^4$ yr, and $t_{cap}$=1.10×$10^4$ yr with non-gravitational forces. Finally, the closest approach to Jupiter is respectively found to be $\bar{d}_{min}$ = 0.19 au (purely gravitational) and $\bar{d}_{min}$ = 0.20 au (including non-gravitational forces).

These values seem to be different from the ones obtained by Fernández & Sosa (2015), where the authors did not report a stable orbit; we therefore assume that they obtained $f_q > 0.2$ and $f_a > 0.1$. This discrepancy could be explained by the different quality of the orbits (now being more accurate) or the integrator algorithm used in both studies. We used a hybrid code that combines a symplectic algorithm with the Burlisch-Stoer code, while Fernández & Sosa (2015) used only Burlisch-Stoer code. Another reason for the discrepancy could be how the clones were built; they generated 50 clones using a Gaussian distribution with a standard deviation given by the osculating uncertainties, while we generated 200 clones, using a Gaussian distribution with a standard deviation given by the $3\times$ osculating uncertainties.

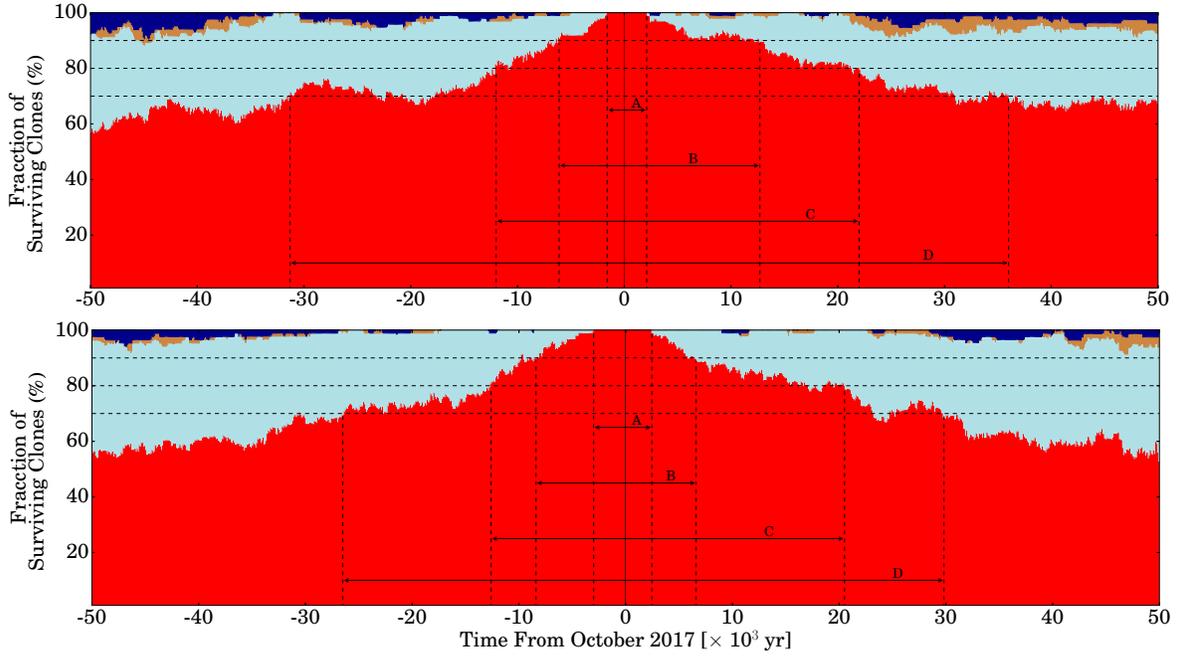

Fig. 12: Statistical orbital evolution of comet 41P and its 200 clones over $10^5$ yr: $5\times10^4$ backward in time and $5\times10^4$ forward. The top panel corresponds to a pure gravitational model, while in the bottom panel non-gravitational forces are included. In both panels different colours refer to different regions in the solar system, namely: red for those in the Jupiter family region, clear blue are Centaurs, yellow are Halley types, and dark blue are trans-Neptunians. Labels A, B, C, and D mean the time spent in the Jupiter family region at different confidence levels, i.e., % of clones in that region. Therefore, A is 100%, B is 90%, C is 80%, and D is 70%.

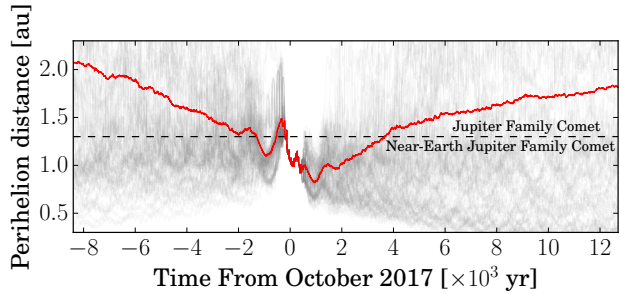

Fig. 13: Perihelion evolution during the time in which comet 41P is in the JFC region with a 90% of CL. Grey orbits correspond with the whole set of clones, which take into account both the pure gravitational model and the model that includes non-gravitational forces. The red line is the average perihelion $\bar{q}$. The dashed horizontal line differentiates between the JFC and NEJFC regions.

Therefore, according to the new results, comet 41P belongs to the *Moderately asteroidal* category (see Fernández & Sosa (2015), their table 3). The authors included in this category the following comets: 197P/LINEAR, 207P/NEAT, 209P/LINEAR, 210P/Christensen, 217P/LINEAR, and 317P/WISE. Therefore, comet 41P is the seventh. All of them generally show little activity compared to normal JFCs. This is in contrast with the comets in the *Highly asteroidal* category, which were reported to have extremely low levels of activity and even stellar-like appearances in some cases. According to Fernández & Sosa (2015), *Moder-*

*ately asteroidal* category still could imply a main asteroid belt origin. However, comet 41P has shown a typical cometary activity based on ice sublimation; therefore, an asteroidal belt origin sounds unlikely. On the other hand, our set of experiments confirms that its orbit is more stable than usual for typical JFCs. Most of them are indeed moving in highly unstable orbits, with capture times in their current near-Earth orbits being lower than a few times $10^3$ yr.

After completing the set of tests proposed by Fernández et al. (2014) and Fernández & Sosa (2015), one wonders about the rough period of time that comet 41P will stay in its current orbit. Using the same sample of 200 clones, we performed a couple of extra experiments to address this question. First, we divided the possible locations of the comet and its clones into four regions depending on their dynamical properties, and we computed the time spent on them, namely:

$$a < \frac{a_S}{(1+e)} \implies \text{JFC type,} \tag{11}$$

$$\left.\begin{array}{c} \frac{a_S}{(1+e)} < a_N \\ e < 0.8 \end{array}\right\} \implies \text{Centaur type,} \tag{12}$$

$$\left.\begin{array}{c} \frac{a_S}{(1+e)} < a_N \\ e > 0.8 \end{array}\right\} \implies \text{Halley type,} \tag{13}$$

$$a > a_N \implies \text{Trans-Neptunian type.} \tag{14}$$

In all these equations, $a_S$ and $a_N$ are the semi-major axes of Saturn and Neptune, respectively. The results of this experiment are displayed in the Fig. 12, where the top panel shows the pure gravitational model and the bottom panel shows the model that includes non-gravitational forces. In this figure, the red area corresponds to JFC types, the clear blue corresponds to Centaur types, the yellow to Halley types and the dark blue to trans-Neptunian types. In particular, we are interested in the amount of time spent in the JFC region, where comets reach a temperature high enough to be periodically active. To extract this information from the figure, we computed the time for different confidence levels (CL), based on the fraction of surviving clones in the JFC region, namely: 100% CL (A); 90% CL (B); 80% CL (C), and 70% CL (D). Therefore, we find that the time spent by 41P in the JFC region is: 3700±100 yr (A); 18800±100 yr (B); 34000±100 yr (C); and 67300±100 yr (D) for the pure gravitational model. For the model that includes non-gravitational forces, we find: 5500±100 yr (A); 15000±100 yr (B); 33100±100 yr (C); and 56300±100 yr (D). In both models we observe that at the end of the simulations, more than half of the clones still remain in the JFC region, with less than 8% of the initial particles being ejected. In the upper panel, we observe that 100% of the particles became JFCs at −1600 ± 100 yr, which matches the result found in Fig. 10 regarding the incursion to its current region after a strong close encounter with Jupiter. In contrast, when we include non-gravitational forces, this incursion occurred earlier. With a 100% CL we obtain that the comet will be in the JFC region for a time ranging between 3700 and 5500 yr. This confirms the stable orbit found in the previous experiments, but favours the trans-Neptunian region as a more plausible origin.

However, 41P is a special case of a JFC in that it belongs to the NEJFC subgroup, that is, a JFC with a perihelion value of $q < 1.3$ au. Our definition of a JFC given in Eq. 11 does not differentiate between JFCs and NEJFCs. Therefore, our second experiment consisted of computing the time the comet was an NEJFC, and focused on the period in which the comet remains in the JFC region, which was from -8400 to 12700 yr, for a 90% CL. Since from both models we obtained very similar results, we decided to compute the average of them. The result is shown in Fig. 13. In the figure, the perihelion evolution of the complete set of clones is displayed in grey, which includes both models, and the average perihelion evolution, $\bar{q}$, shown in red. We divided the physical space between the JFC and NEJFC regions. We observe that the dynamical evolution will carry 41P to a minimum perihelion distance of $q = 0.82$ au 950 yr from now. The evolution of the clones is extremely compact during a period of 750 yr around the current time (from -300 to 450 yr), the equivalent to ∼ 140 orbits. During this period, there is no divergence in the orbits. From the current epoch until 3600 yr, $\bar{q} < 1.3$ au, therefore we considered this time as the period in which the comet 41P will be in Earth's neighbourhood. All these results suggest that the orbit of 41P is more stable than typical JFCs. The time in the NEJFC region will last roughly 3600 years, which is equivalent to ∼665 orbits.

## 5. Summary & Conclusions

In the first part of this work, we describe and analyse an extensive observational data set of images of comet 41P obtained with the TRAPPIST telescopes using dust tail models. Those observations cover the main portion of the orbital arc in which the comet displayed activity, from 1.5 au pre-perihelion to 1.7 au post-perihelion. In our dust models we followed the assumptions and guidelines given by Moreno et al. (2017a) for the study of comet 67P, which was the Rosetta target.

Our main conclusion is that it is not possible to explain the complete set of observations using a full isotropic ejection model. In fact, we find that a complex ejection pattern which switched from full isotropic to anisotropic (February 24-March 14), and then back from anisotropic to full isotropic again on June 7-28 provides the best description of the observations. During the anisotropic period, we find that ∼ 90% of the ejected particles came from two strongly active areas, one located in the northern hemisphere and the other in the southern. This model is in agreement with the recent discovery of the fast rotational period variation reported by Bodewits et al. (2018) from March to May, 2017, in the sense that the two powerful active areas could have acted as brakes, increasing the nucleus rotation period. However, the location found in our model for these active areas prevents us from giving final confirmation, and leads us to the consideration that other factors may be acting and possibly affecting the fast spin-down observed. Further investigations are encouraged.

In general terms, from the dust model we obtain that the total dust mass ejected is ∼ $7.5 \times 10^8$ kg. This quantity is roughly the total dust ejected by the comet during the whole orbit. This amount of dust is low compared to other comets of the same family; however, 41P is a small comet, and this quantity represents a non-negligible fraction of its total mass. This implies that 41P suffered a substantial amount of erosion during its last incursion to perihelion. From observations of gases also performed with TRAPPIST telescopes (Moulane et al. 2018), we find that the dust-to-water mass ratio is low ranging from 0.25 to 1.5. The complete set of dust parameters, which best describe the evolution of its dust environment, is also reported, which includes the maximum particle size, the power-law index of the size distribution, and the ejection velocity field of the particles. All these results confirm the dust-poor nature of 41P.

In the second part of this work, we explored the dynamical nature of 41P with numerical simulations. We followed the set of tests proposed by Fernández et al. (2014) and Fernández & Sosa (2015) to constrain the degree of instability of its orbit. In those experiments, we always considered two models: a pure gravitational model and a model that included non-gravitational forces. No significant differences between them were found. We obtain that 41P is more stable than typical JFCs, its orbit being in the category *Moderately asteroidal*, which could imply a main asteroid belt origin. However, the complete set of dynamical experiments performed, and the activity being driven by ice-sublimation, favour the trans-Neptunian origin hypothesis. The status of NEJFC, that is, a JFCs with a perihelion distance of $q < 1.3$ au, seems to be relatively new for this comet. The expected period of time during which the comet will remain in this region is roughly ∼ 3600 yr. A minimum perihelion distance will be reached in 950 years, with a value of $q \sim 0.8$ au.

With the information currently available, we estimate the total mass of the comet to be $1.6 \times 10^{12}$ kg. If the dust production rate per orbit of ∼ $7.5 \times 10^8$ kg remains constant or similar, every incursion to perihelion during which the comet is in Earth's vicinity will mean a prolonged period of nucleus erosion, and it may lose up to 30% of its mass within 3600 yr; even more if we also consider the gas production rates. This erosion could be even larger since the perihelion distance will decrease by ∼0.2 au over the next 950 years. This fact, combined with its low dust-to-gas mass ratio, and its propensity to undergo outbursts (Kresak 1974; Bodewits et al. 2018) could provoke the disruption of the nucleus in a relatively short period of time.

*Acknowledgements.* This work is supported by a Marie Curie CO-FUND fellowship, co-founded by the University of Liège and the European Union. TRAPPIST-South is funded by the Belgian Fund for Scientific Research (Fond National de la Recherche Scientifique, FNRS) under the grant FRFC 2.5.594.09.F, with the participation of the Swiss FNS. TRAPPIST-North is a project funded by the University of Liège, and performed in collaboration with Cadi Ayyad University of Marrakesh. E.J. is a Belgian FNRS Senior Research Associate, and M.G. is an FNRS Research Associate. We thank the anonymous referee for having significantly improved the paper and Fernando Moreno and Damien Hutsemékers for their valuable discussions. Some of the simulations in this paper made use of the REBOUND code which can be downloaded freely at http://github.com/hannorein/rebound. We thank Hanno Rein and Daniel Tamayo for their help using REBOUND.